\documentclass[twocolumn]{aastex63}
\usepackage{amssymb,amsmath}
\usepackage{graphicx}
\usepackage{xcolor,natbib}
\usepackage{multirow}
\usepackage[T1]{fontenc}
\usepackage{ae,aecompl}
\usepackage{newtxtext, newtxmath}
\usepackage{float}
\usepackage{subfigure}
\usepackage{longtable}
\usepackage{enumitem}
\usepackage{longtable,listings}
\usepackage[flushleft]{threeparttable}
\usepackage{parcolumns}
\usepackage{makecell}

\def\oc3{[O~{\sc iii}]$_c$}
\def\ob3{[O~{\sc iii}]$_b$}
\def\dif{\mathop{}\!\mathrm{d}}

\accepted{\today, ApJ}
\shorttitle{Effects of AGN variability on QPO}
\shortauthors{ZHANG}

\begin{document}

\title{Effects of intrinsic AGN variability on optical QPOs related to sub-pc binary black hole systems in broad line AGN}

\correspondingauthor{XueGuang Zhang}% \email{xgzhang@gxu.edu.cn}}
\email{xgzhang@gxu.edu.cn}
\author{XueGuang Zhang$^{*}$}
\affiliation{Guangxi Key Laboratory for Relativistic Astrophysics, School of Physical Science and Technology, GuangXi 
University, Nanning, 530004, P. R. China}

\begin{abstract} %%%about 178 words
	In this manuscript, an oversimplified model is proposed to test the effects of intrinsic AGN variability on expected 
optical quasi-periodic oscillations (QPOs) related to sub-pc binary black hole systems (BBHs) in broad line AGN. The commonly 
accepted CAR (Continuous AutoRegressive) process is applied to describe intrinsic AGN variability related to each BH accreting 
system in a sub-pc BBH system. Considering obscurations related to orbital rotations with periodicity $T_p$, artificial light 
curves including signals for optical QPOs can be built. Then, comparing the intrinsic periodicities $T_p$ with the measured 
robust periodicities $T_o$ through the artificial light curves, distributions of $T_p/T_o$ have four significant peaks around 
1, 2, 3 and 4, leading less than half of the artificial light curves to have consistency between $T_o$ and $T_p$. Moreover, 
different collected model parameters have few effects on the distributions of $T_p/T_o$, indicating the effects of intrinsic 
AGN variability on optical QPOs are significantly strong and stable. Furthermore, after checking properties of optical QPOs in 
the light curves with different time steps, there are tiny effects of time steps on optical QPOs.  
\end{abstract}


\section{Introduction}

%%%1st
	Sub-pc binary black hole systems (BBHs) are the preferred carriers on studying hierarchical formation and evolution 
of black holes. Due to observational technique limitations on spatially resolved photometric images, the long-standing optical 
quasi-periodic oscillations (QPOs) with periodicities around hundreds to thousands days have been accepted as good indicators 
for sub-pc BBHs in broad line active galactic nuclei (AGN). There are so far more than 200 candidates for ub-pc BBHs through 
implications of optical QPOs, such as the reported 1800days optical QPOs in PG 1302-102\ in \citet{gd15a, kp19}, the 540days 
optical QPOs in PSO J334.2028+01.4075\ in \citet{lg15}, the optical QPOs in two samples of more than 160 optical QPOs in 
\citet{gd15, cb16}, the 1500days optical QPOs in SDSS J0159 in \citet{zb16}, the 1150days optical QPOs in Mrk915 in \citet{ss20}, 
the 1.2yr optical QPOs in Mrk231 in \citet{ky20}, the 1607days optical QPOs in SDSS J0252 in \citet{lw21}, the 6.4yr optical 
QPOs in SDSS J0752 in \citet{zh22a}, the 3.8yr optical QPOs in SDSS J1321 in \citet{zh22c}, the 340days optical QPOs in SDSS 
J1609 in \citet{zh23a}, the 1000days optical QPOs in SDSS J1257 in \citet{zh23b} and the 550days optical QPOs in the known 
PG 1411+442 in \citet{zh25}, etc. Here, the optical QPOs and the corresponding periodicities discussed in the 
manuscript are related to sub-pc BBH systems, not related to transient behavior \citep{cz98, ws01, sm18, gt19} arising in 
the accretion disk around single supermassive BH.

%%2nd
	Although the optical QPOs can be well applied to detect sub-pc BBHs in broad line AGN, considering intrinsic AGN 
variability as one of the fundamental characteristics of AGN \citep{um97, gh07, wm09, sl19, zh23c}, it is not clear for 
effects of intrinsic AGN variability on the detected optical QPOs in broad line AGN. Whether the effects of can lead to 
detected periodicities very different from the intrinsic values? To discuss the question is our main objective of this 
manuscript. Moreover, due to sub-pc BBHs as strongly preferred candidates for sources of nanohertz frequency gravitational 
waves as discussed in \citet{sr15, th16, ml17, sh18, aa19, ab23, ct24}, to report clear effects of intrinsic AGN variability 
on optical QPOs related to sub-pc BBHs in broad line AGN can provide more confident results related to GWs produced by 
sub-pc BBHs.

%%%3rd
	It has been accepted that the CAR (Continuous AutoRegressive) process \citep{kbs09} and/or DRW (damped random walk) 
process \citep{koz10, zk13, zk16} can be applied to describe long-term intrinsic AGN variability \citep{mi10}. Certainly, as 
shown in \citet{mi10}, there are about $\sim$2\% of the quasars of which light curves cannot be well described by the CAR 
process. Moreover, as discussed in \citet{kb14, kv17, mv19, yr22, kg24}, rather than the simple CAR process,
higher order CARMA(p, q) ($p\ge1$ and $q\le p$) (Continuous AutoRegressive Moving Average) process should be possibly preferred 
to describe AGN variability. However, considering the smaller number ratio $\sim$2\% of quasars of which light curves cannot 
be well descried by the CAR process, and considering well-determined distributions of the CAR process parameters 
in common quasars variability but none-determined distributions of parameters in the CARMA process, the CAR process is 
accepted to widely describe intrinsic AGN variability. Meanwhile, considering the high quality light 
curves from the sky survey project of Zwicky Transient Facility (ZTF) \citep{bk19, ds20}, simulated light curves with 
similar time information from ZTF can be created by the CAR process. Therefore, combinations of obscurations due to orbital 
rotations as described in Section 2, effects of intrinsic AGN variability can be determined on expected optical QPOs related 
to sub-pc BBHs. This manuscript is organized as follows. Section 2 presents our model and main hypotheses, our main 
results and necessary discussions. Section 3 gives our main conclusions. And we have adopted the cosmological parameters of 
$H_{0}=70{\rm km\cdot s}^{-1}{\rm Mpc}^{-1}$, $\Omega_{\Lambda}=0.7$ and $\Omega_{\rm m}=0.3$ throughout this manuscript.

%%%figure1_exams.pro
\begin{figure*}
\centering\includegraphics[width = 18cm,height=18cm]{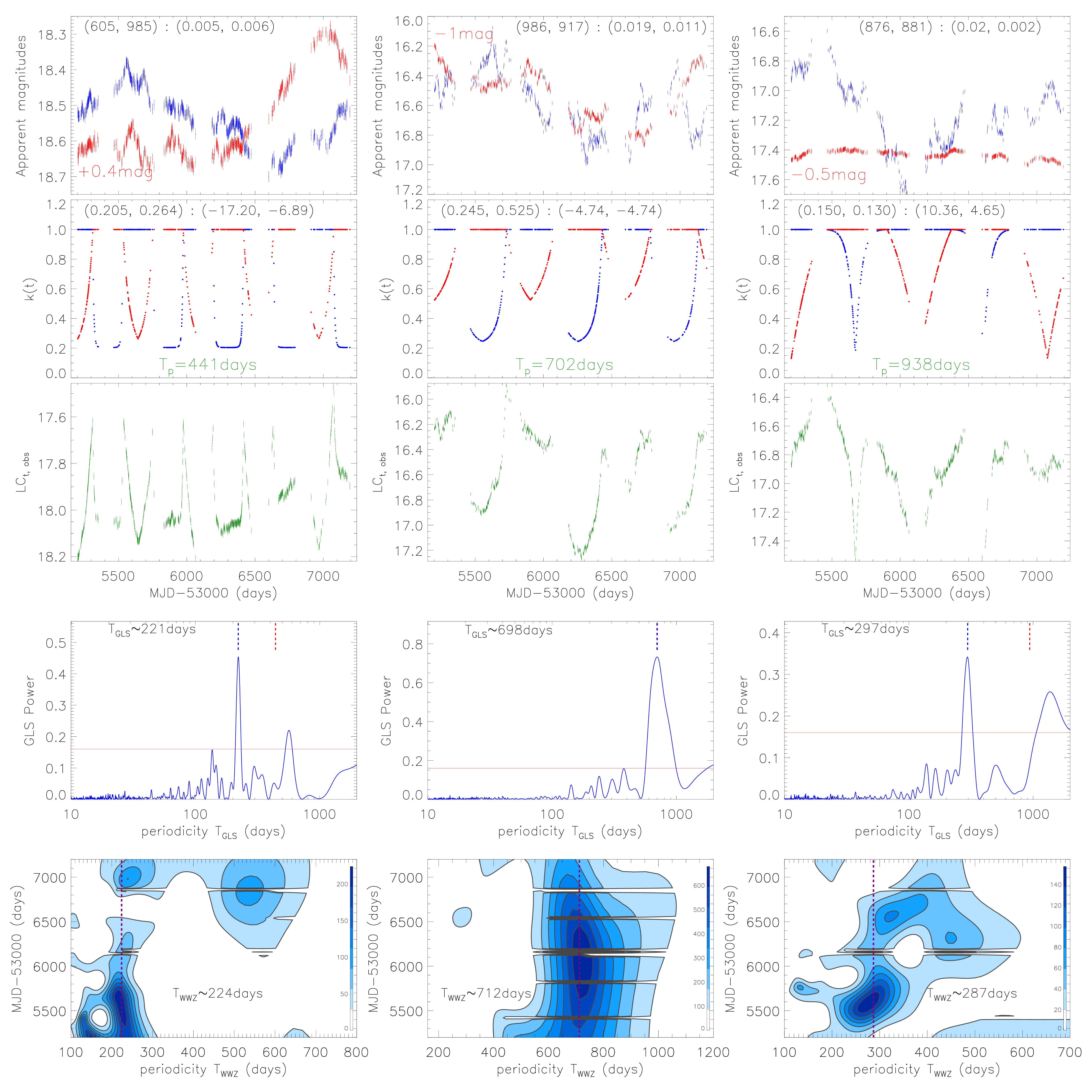}
\caption{Top panels show the CAR process created $LC_{t,~1}$ (in blue) and $LC_{t,~2}$ (in red) by different process parameters 
of ($\tau_1$,~$\tau_2$) (in units of days) : ($\sigma_{*,1}$,~$\sigma_{*,2}$) (in units of $mag/days^{0.5}$) shown in the top 
region of each panel. The panels in the second row show the applied $k_1(t)$ (in blue) and $k_2(t)$ (in red) by different 
parameters of ($k_{10}$,~$k_{20}$) : ($\alpha_1$,~$\alpha_2$) shown in the top region of each panel and by different orbital 
periodicity $T_p$ marked in dark green characters in bottom region of each panel. The panels in the third row show the
determined $LC_{t,~obs}$ after considering the obscurations due to orbital rotations. The panels in the fourth row show the GLS 
determined power properties of the corresponding $LC_{t,~obs}$, with the horizontal red line marking the 5$\sigma$ confidence 
levels based on white noise simulations. In each panel in the fourth row, the vertical dashed blue line marks the 
GLS determined periodicity $T_o$ (=$T_{GLS}$), and the vertical dashed red line marks the position for the input $T_p$. 
The bottom panels show the WWZ determined two dimensional power properties of the corresponding $LC_{t,~obs}$, 
with frequency step of 0.0001 and searching periodicities from 100 days to 2000 days. In each bottom panel, vertical dashed 
purple line marks the position of the WWZ determined periodicity $T_{WWZ}$ which has been listed in the bottom region of each 
bottom panel. In each bottom panel, as the shown colorbar in the right hand, the contour levels in different colors show the 
numbers of data points in the evenly divided 1990$\times$100 regions in the shown space of $T_{WWZ}$ versus MJD-53000.}
\label{exam}
\end{figure*}

%%%figure_wwz.pro
\begin{figure}
\centering\includegraphics[width = 8cm,height=5.5cm]{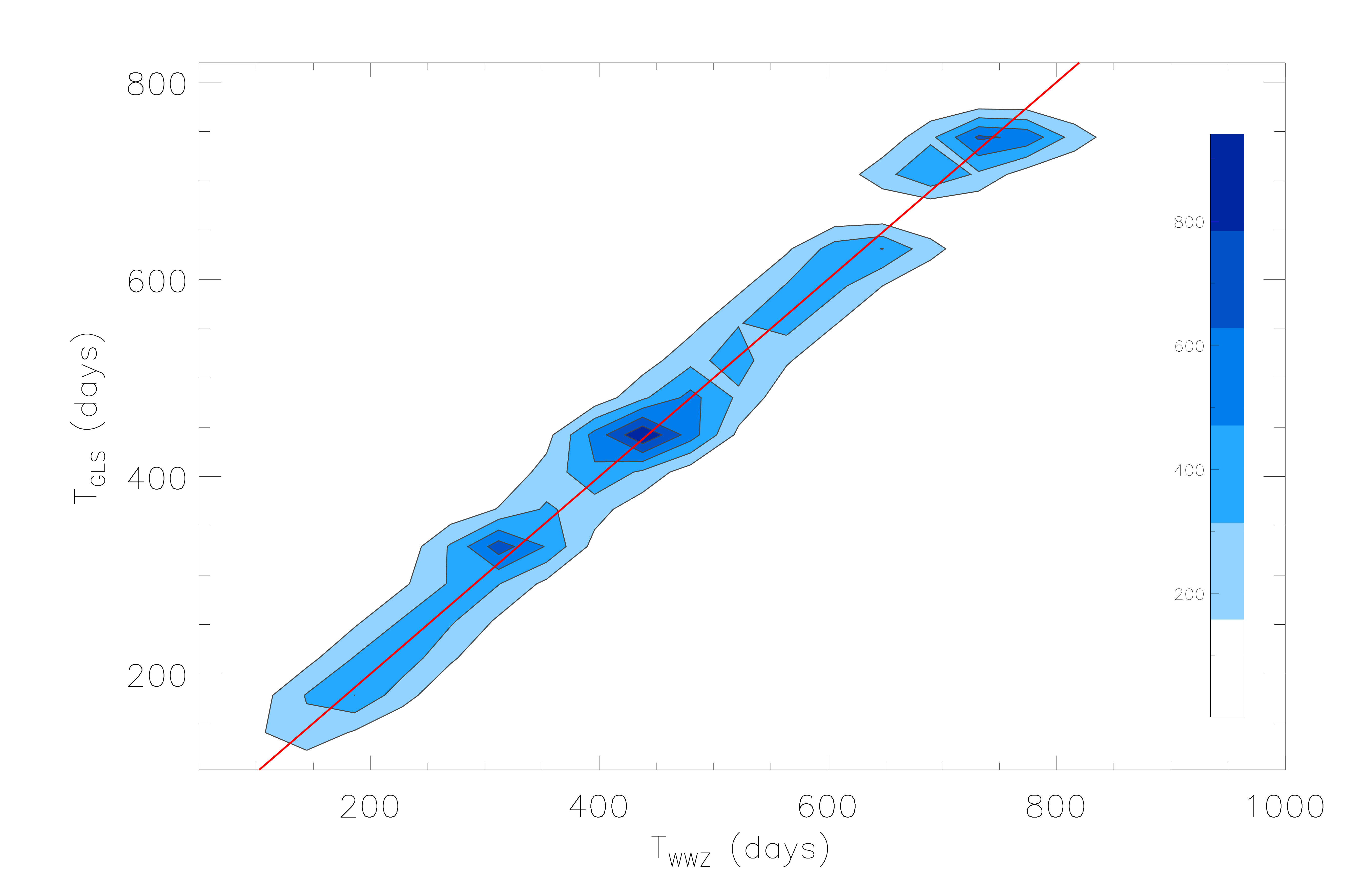}
\caption{On the correlation between $T_{GLS}$ and $T_{WWZ}$ for the artificial light curves. The solid red line 
shows $T_{GLS}=T_{WWZ}$. As the shown colorbar in the right hand, the contour levels in different colors show the numbers 
of data points in the evenly divided 25$\times$25 regions in the shown space of $T_{WWZ}$ versus $T_{GLS}$.
}
\label{wwz}
\end{figure}          

%%%recheck_rednoise.pro
\begin{figure}
\centering\includegraphics[width = 8cm,height=5.5cm]{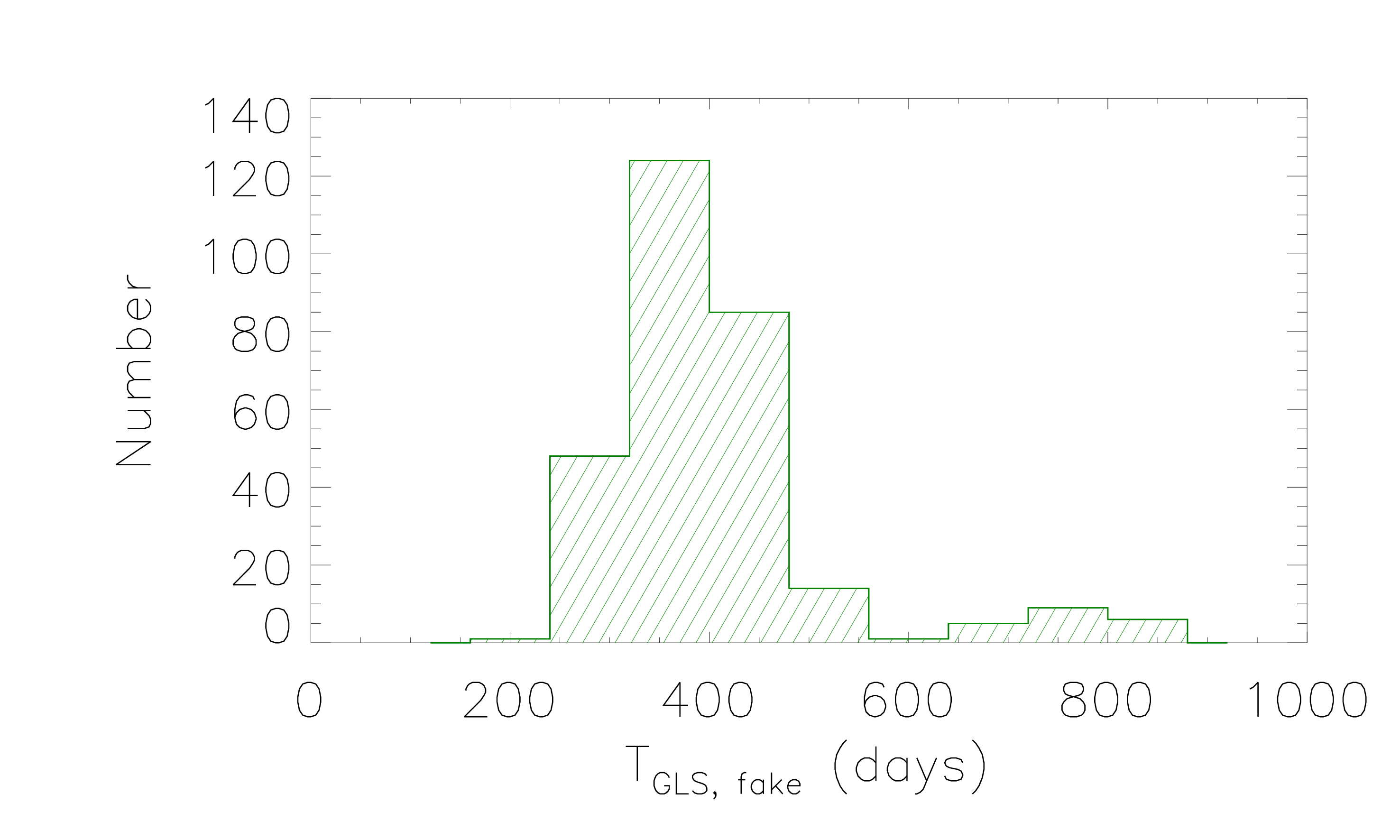}
\caption{Distributions of the 293 $T_{GLS, fake}$ determined by applications of the GLS method in the 293 
of the 100000 artificial light curves created by CAR process.}
\label{red}
\end{figure}

%%%figure1_exams.pro
\begin{figure*}
\centering\includegraphics[width = 18cm,height=6cm]{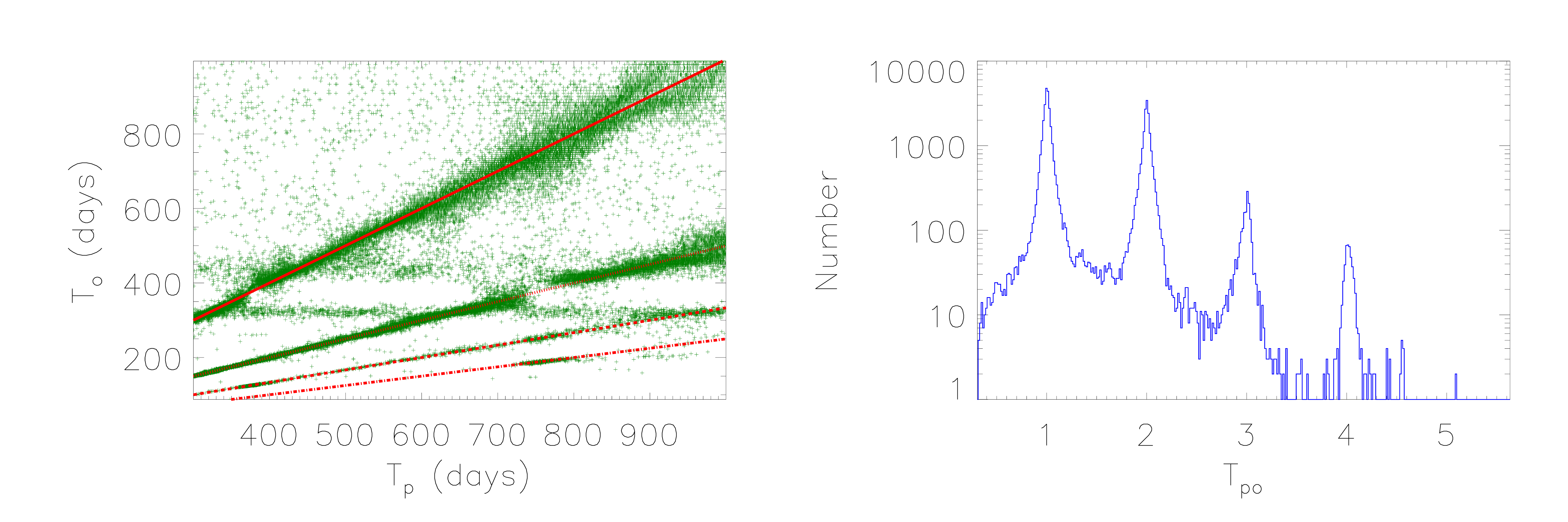}
\caption{Left panel shows the correlation between the input periodicity $T_p$ and the measured periodicity $T_o$ from the  
$LC_{t,~obs}$ including obscurations due to orbital rotations. Solid red line, dotted red line, dashed red line and dot-dashed 
red line show $T_p=T_o$, $T_p=2\times T_o$, $T_p=3\times T_o$ and $T_p=4\times T_o$, respectively. Right panel shows the 
distribution of $T_{po}$ (ratio of $T_p$ to $T_o$). In right panel, in order to show clearer features of the distribution, 
the plot in y-axis is shown in logarithmic coordinate.}
\label{tpo}
\end{figure*}

\section{Main Results and necessary Discussions} 

\subsection{Model and main hypotheses}

%%%1st
	In order to describe the expected periodic photometric variability properties of an assumed sub-pc BBH system by 
periodic variations of obscurations due to orbital rotations, the following three steps are applied in this manuscript.

%%%2nd
	First, totally accepted the CAR process discussed in \citet{kbs09}, two artificial long-term light curves (in photometric 
apparent magnitudes) $LC_{t,~1}$ and $LC_{t,~2}$ from two BH accreting systems in an assumed sub-pc BBH system can be created by
\begin{equation}
\begin{split}
\dif LC_{t,~1}&=\frac{-1}{\tau_1}LC_{t,1}\dif t+\sigma_{*,2}\sqrt{\dif t}\epsilon(t,1)~+~LC_{m1} \\
\dif LC_{t,~2}&=\frac{-1}{\tau_2}LC_{t,2}\dif t+\sigma_{*,1}\sqrt{\dif t}\epsilon(t,2)~+~LC_{m2}
\end{split}
\end{equation}
Here, there are no considerations of any physical mechanisms on probable formations and/or evolutions of sub-pc 
BBHs, but only considering the simple point that two BH accreting systems in an assumed sub-pc BBH system can lead to 
two sources of photometric variability. Moreover, the main objective of the manuscript is to check effects of intrinsic AGN 
variability on expected optical QPOs related to sub-pc BBHs. Therefore, in an assumed sub-pc BBH system, we have accepted 
that each BH accreting system has its optical emission regions to be independent and complete. Otherwise, the mixed 
optical emission regions of the two BH accreting systems could not lead to expected optical QPOs. Therefore, the CAR process 
can be applied to trace the variability of the two BH accreting systems, and effects of sizes of accretion disks are not 
discussed any more in this manuscript. Furthermore, the BBHs on sub-pc scales harboring two individual BH accreting 
systems are mainly considered in the manuscript, not binary black hole systems embedded in the AGN gaseous disks as more 
recent discussions in \citet{ig20, fs24, le25}. In other words, considering the two BH accreting systems in sub-pc BBHs 
we mainly focused on have separated distance larger enough, even there were probable circumbinary disks around sub-pc BBH 
systems, there were few contributions of such circumbinary disks to optical emissions. Therefore, there are no further 
discussions on effects of probable circumbinary disks on our following results.

%%3rd
	Here, considering high quality light curves from the ZTF, the time information $t$ for the $LC_{t,~1}$ and $LC_{t,~2}$ 
are similar as the time information of SDSS quasars in ZTF. Here, the time information is collected from the ZTF g-band light 
curve of PG 1411+442 of which optical QPOs have been discussed in our more recent paper \citet{zh25}, a known reverberation 
mapped broad line AGN in \citet{ka00, pe04}. Certainly, there are different time durations and different numbers of data points 
for different quasars in ZTF, however, there are few effects of time information on our final results, unless there are vary 
small number of data points in the ZTF light curve of the collected quasar.

%%4th
	Meanwhile, the two process parameters of intrinsic variability timescales $\tau_1$ and $\tau_2$ in the CAR process are 
randomly collected from 100days to 1000days, the common values in quasars as discussed in \citet{kbs09, mi10}. And, rather than 
giving values of $\sigma_*$, the parameters of $\frac{\tau_1\times~\sigma_{*,1}^2}{2}$ and $\frac{\tau_2\times~\sigma_{*,2}^2}{2}$ 
are randomly collected from 0.001 to 0.32, representing variances of $LC_{t,~1}$ and $LC_{t,~2}$. The values from 0.001 to 0.32 
are the common values of variances of ZTF light curves of SDSS quasars. And the parameters of $LC_{m1}$ and $LC_{m2}$, as mean 
apparent magnitudes of $LC_{t,~1}$ and $LC_{t,~2}$, are randomly collected from 16 to 19, common values for low redshift SDSS 
quasars.

%%%5th
	Moreover, based on the uncertainties $LC_{t,~err,~pg}$ of the ZTF g-band light curve $LC_{t,~PG}$ of PG 1411+442, 
the uncertainties $LC_{t,~err,~1}$ and $LC_{t,~err,~1}$ of the $LC_{t,~1}$ and $LC_{t,~2}$ can be simply determined as 
\begin{equation}
\frac{LC_{t,~err,~1}}{LC_{t,~1}} = \frac{LC_{t,~err,~pg}}{LC_{t,~PG}} = \frac{LC_{t,~err,~2}}{LC_{t,~2}}
\end{equation}.
Top panels of Fig.~\ref{exam} shows three examples on $LC_{t,~1}$ and $LC_{t,~2}$ and the corresponding uncertainties.

%%%6th
	Second, based on the created artificial light curves of $LC_{t,~1}$ and $LC_{t,~2}$, there are two periodic varying 
parameters $k_1(t)$ and $k_2(t)$ applied to trace obscurations due to orbital rotations with randomly selected periodicity 
$T_p$ larger than 300days but smaller than 1000days, leading to the observed light curve $LC_{t,~obs}$ as 
\begin{equation}
	10^{\frac{LC_{t,~obs}}{-2.5}} = k_1(t)~\times~10^{\frac{LC_{t,~1}}{-2.5}} + 
	k_2(t)~\times~10^{\frac{LC_{t,~2}}{-2.5}}
\end{equation}.
Here, $k_1(t)$ and $k_2(t)$ with $t$ within each cycle are created by piecewise functions as
\begin{equation}
k_1(t) = 
\begin{cases}
	1 & (0\le\frac{t-t_0}{T_p}<\frac{1}{4}) \\
	A_1+B_1~t^{\alpha_1} & (\frac{1}{4}\le\frac{t-t_0}{T_p}<\frac{1}{2}) \\
	A_1+B_1~(T_p~-~t)^{\alpha_1} & (\frac{1}{2}\le\frac{t-t_0}{T_p}<\frac{3}{4}) \\
	1 & (\frac{3}{4}\le\frac{t-t_0}{T_p}\le1)
\end{cases}
\end{equation}
\begin{equation}
k_2(t) = 
\begin{cases}
	A_2+B_2~t^{\alpha_2} &  (0\le\frac{t-t_0}{T_p}<\frac{1}{4}) \\
	1 & (\frac{1}{4}\le\frac{t-t_0}{T_p}<\frac{1}{2}) \\
	1 & (\frac{1}{2}\le\frac{t-t_0}{T_p}<\frac{3}{4}) \\
	A_2+B_2~(T_p~-~t)^{\alpha_2} & (\frac{3}{4}\le\frac{t-t_0}{T_p}\le1)
\end{cases}
\end{equation}
Here, $k_1(t)$ and $k_2(t)$ have 1 as the maximum values, indicating none obscuration effects, and have $0\le k_{10}\le0.6$ 
and $0\le k_{20}\le0.6$ as the minimum values, indicating apparent obscuration effects due to orbital rotations. And $k_{10}=0$ 
and $k_{10}>0$ ($k_{20}=0$ and $k_{20}>0$) mean totally obscured and partly obscured. Simple discussions on different upper 
boundaries of minimum values of $k_{10}$ and $k_{20}$ from 0.6 can be found in the following subsection. And due to the time 
duration about 2000days of the artificial light curves, the selected periodicity is not larger than 1000days, to confirm the 
time durations at least two times larger than the periodicity, and the selected periodicity is not smaller than 300days after 
considering the minimum periodicity about 300days in the sample of QPOs in \citet{gd15}. Once the input parameters 
$k_{10}$, $\alpha_1$, $k_{20}$, $\alpha_2$ and the input periodicity $T_p$ are clearly given, then the factors of $A$ and $B$ 
listed in the equation (4) and equation (5) can be clearly determined. The panels in the second row of Fig.~\ref{exam} show 
three examples of $k_1(t)$ and $k_2(t)$ applied to the light curves shown in the top panels of Fig.~\ref{exam}, and the 
corresponding observed light curve $LC_{t,~obs}$ after considering the orbital rotations expected obscurations traced by 
$k_1(t)$ and $k_2(t)$. Actually, different values of $\alpha_1$ from $\alpha_2$ and $k_{10}$ from $k_{20}$ can be reasonably 
accepted due to probably different properties of physical and geometric structures of the central two BH accreting systems. 
%Effects of different values of $\alpha_1$ from $\alpha_2$ and 
%$k_{10}$ from $k_{20}$ will be discussed in the next section.

%%%7th
	Here, we should note that the periodic variations of obscuration is the key role leading to expected optical QPOs in 
an assumed sub-pc BBH system. However, it is hard to build an efficient model, not only considering physical properties of the 
materials for obscurations but also considering spatial properties of the materials. Therefore, we proposed the piecewise 
functions, equation (4) and equation (5), which can be applied to simply describe periodic variations of the obscurations 
expected by an assumed sub-pc BBH system. The different values of $\alpha_1$ and $\alpha_2$ can be applied to describe the 
corresponding properties of obscurations related to different properties of materials for obscurations. Meanwhile, for an 
assumed sub-pc BBH system, it is necessary to consider inclination of the orbit with respect to the line of sight. The 
inclination could lead to one BH accreting system only partly obscured by the other BH accreting system, indicating larger 
values of $k_{10}$ and $k_{20}$ applied in the piecewise functions. Therefore, different values of $k_{10}$, $\alpha_1$ and 
$k_{20}$, $\alpha_2$ can be applied to simply consider the effects of materials source for the obscurations and also the 
effects of inclination of the orbit.

%%%8th
	Third, for each artificial light curve $LC_{t,~obs}$, the widely accepted generalized Lomb-Scargle (GLS) method 
\citep{ln76, sj82, zk09, vj18} is applied to check periodicity. The panels in the fourth row of Fig.~\ref{exam} show the GLS 
method determined power properties for the corresponding $LC_{t,~obs}$ shown in the panels in the third row of Fig.~\ref{exam}. 
Meanwhile, in order to test robustness of the detected periodicity by the GLS method, the known WWZ (Weighted 
wavelet z-transform) technique \citep{fg96, al16, gt18, ks20} has been applied to re-determine the periodicity in each 
artificial light curve, similar as what we have recently done in \citet{zh23a, zh25}. The bottom panels of Fig.~\ref{wwz} 
shows the determined two-dimensional WWZ power properties for the artificial light curves shown in the panels in the third 
row of the Fig.~\ref{exam}, leading to the WWZ determined periodicity $T_{WWZ}$ consistent with the GLS determined periodicity 
$T_{GLS}$ in the three shown artificial light curves as examples.

%%%9th
	Therefore, a simple procedure can be applied to create artificial light curves $LC_{t,~obs}$ including apparent 
obscurations due to orbital rotations, based on the three steps above including 11 parameters, the CAR process parameters 
of $\tau_1$, $\tau_2$, $\sigma_{*,1}$, $\sigma_{*,2}$, mean magnitudes $LC_{m1}$ and $LC_{m2}$ of light curves $LC_{t,~1}$ 
and $LC_{t,~2}$, minimum values of $k_{10}$ and $k_{20}$, input periodicity $T_p$, power indices of $\alpha_1$ and 
$\alpha_2$. Finally, based on the randomly selected 11 parameters, 50000 artificial light curves $LC_{t, obs}$ are created, 
and the corresponding periodicities $T_{GLS}$ and $T_{WWZ}$ are determined by the GLS method and the WWZ technique.

%%%10th
	Before ending the section, three additional points are noted. Fist and foremost, Fig.~\ref{wwz} shows the correlation 
between the determined periodicity $T_{GLS}$ by the GLS method and the determined periodicity $T_{WWZ}$ by the WWZ technique 
for the artificial light curves. There is a strong linear correlation between $T_{GLS}$ and $T_{WWZ}$ with Spearman Rank 
correlation coefficient about 0.91 ($P_{null}~<~10^{-10}$), and with the mean value of $T_{GLS}/T_{WWZ}$ about $0.998\pm0.081$ 
(the mean value plus/minus the standard deviation). The results in Fig.~\ref{wwz} strongly indicate that the measured 
periodicities in the artificial light curves to be robust enough. After considering the linear correlation between $T_{GLS}$ 
and $T_{WWZ}$, the GLS determined periodicities $T_{GLS}$ are accepted as the measured periodicity $T_{o}$ in the artificial 
light curves. Actually, applications of $T_{WWZ}$ as $T_{o}$ can lead to the similar final results. 

%%11th
	Besides, we do focus on the expected optical QPOs after main considerations of obscurations due to orbital rotations 
which can be described by piecewise functions described by equation (3), equation (4) and equation(5), not on the subtle 
spatial structures. In other words, rather than applications of physical mechanisms, a mathematics-oriented approach is 
applied, leading to the similar photometric variability properties expected by the physical mechanism related to the sub-pc 
BBHs. 

%%%12th
	Last but not the least, although the mathematics-oriented approach above can lead to expected optical QPOs, it is 
also necessary to check the probability of optical QPOs actually from intrinsic AGN variability, i.e., to check the effects 
of red nosies on the detected QPOs. Here, similar as what we have recently done in \citet{zh23a, zh25} and similar as 
discussed results in \citet{vu16}, fake QPOs related to red noises can be checked in the 100000 artificial light curves 
(50000 $LC_{t,1}$ and 50000 $LC_{t,2}$) which have been created by equation (1) above. Accepted the simple criterion that 
there is significant peak within the range from 100days to 1000days in the GLS power above the $5\sigma$ confidence level 
through white noise simulations, QPOs with GLS method determined periodicities $T_{GLS, fake}$ can be detected in 293 of 
the 100000 artificial light curves, leading to the probability\footnote{The probability is about 6 times larger than the 
probability $4.8\times10^{-4}$ in \citet{zh25}. However, considering that there are no restrictions on CAR parameters no 
restrictions on range of periodicity, the larger probability can be reasonably accepted.} about $293/100000=2.93\times10^{-3}$ 
for detected fake QPOs related to red noises. Distribution of the 293 $T_{GLS, fake}$ is shown in Fig.~\ref{red}. Therefore, 
there are few effects of red noises on detecting QPOs. Moreover, in the procedure above, pointed periodic variations of 
obscurations can apparently lead to expected periodic signals in the artificial light curves $LC_{t, obs}$, thus only 
simple discussions above on effects of red noises are given, and there are no further discussions on effects of red noises.

\subsection{Main Results}

%%1st
	It is necessary to check whether are there the consistency between the input periodicity $T_p$ and the measured 
periodicity $T_{o}$ for the artificial light curves. Left panel of Fig.~\ref{tpo} shows the correlation between $T_p$ and 
$T_{o}$ for the 50000 $LC_{t, obs}$. Right panel of Fig.~\ref{tpo} shows the distribution of the ratio $T_{po}$ of $T_p$ 
to $T_{o}$. It is clear that part of the artificial light curves $LC_{t, obs}$ have the measured periodicity $T_{o}$ to be 
two to four times smaller than the input periodicity $T_p$. Meanwhile, as the shown examples in Fig.~\ref{exam}, the 
corresponding $T_{po}$ can be found as $T_{po}\sim1.99$, $T_{po}\sim1.01$ and $T_{po}\sim3.2$, respectively. And among the 
50000 $LC_{t, obs}$, there are $N_{po1}=29240$, $N_{po2}=18336$, $N_{po3}=1901$ and $N_{po4}=523$ as the numbers of 
$LC_{t, obs}$ with $T_{po}\le1.5$, with $1.5\le T_{po}\le2.5$, with $2.5\le T_{po}\le3.5$ and with $T_{po}\ge3.5$. The 
number rations are about $N_p=N_{po1}:N_{po2}:N_{po3}:N_{po4}\sim56:35:3.6:1$. Therefore, about 41.52\% 
($(N_{po2} + N_{po3} + N_{po4})/50000$) of the 50000 $LC_{t, obs}$ have the measured periodicity $T_{o}$ very smaller (at 
least two times smaller) than the input periodicity $T_p$, due to the expected apparent effects of intrinsic AGN variability. 
In detail, about 36.67\% ($N_{po2}/50000$), 3.8\% ($N_{po3}/50000$) and 1.05\% ($N_{po4}/50000$) of the 50000 $LC_{t, obs}$ 
have the measured periodicity $T_{o}$ about two times smaller, three times smaller and at lest four times smaller than the 
input periodicity $T_p$, respectively. Here, through the linear correlations shown in the left panel of Fig.~\ref{tpo}, 
the GLS determined periodicity can be reasonably accepted as the measured periodicity $T_o$ of the QPOs in the simulated 
$LC_{t,obs}$ with few effects of red noises, otherwise there were no correlations between the measured $T_o$ and the input 
$T_p$.

%%%figure1_exams.pro
\begin{figure}
\centering\includegraphics[width = 8cm,height=22cm]{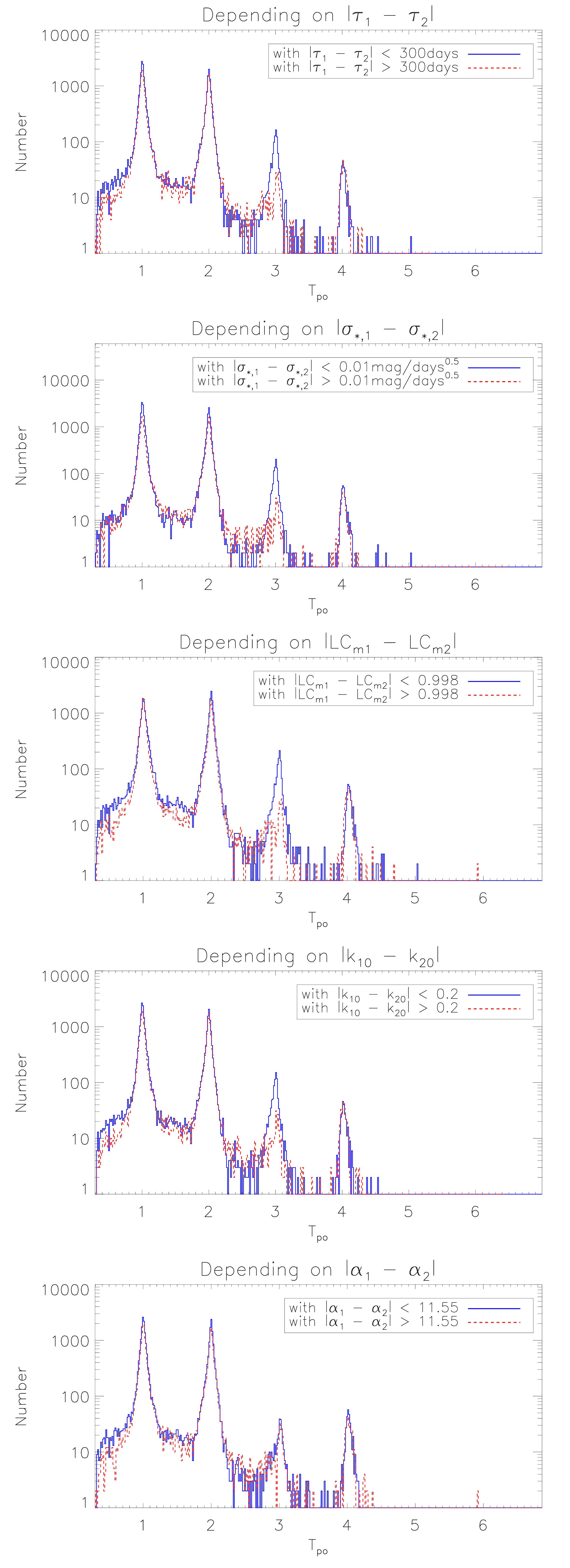}
\caption{In each panel, distributions of $T_{po}$ are compared for the $LC_{t, obs}$ with the different ranges of the 
difference of the model parameter pairs. The applied model parameter pair are listed in title of each panel, the mean value 
of the difference of the model parameter pair is marked in top right region of each panel. }
\label{mpo}
\end{figure}

\begin{figure}
\centering\includegraphics[width = 8cm,height=5cm]{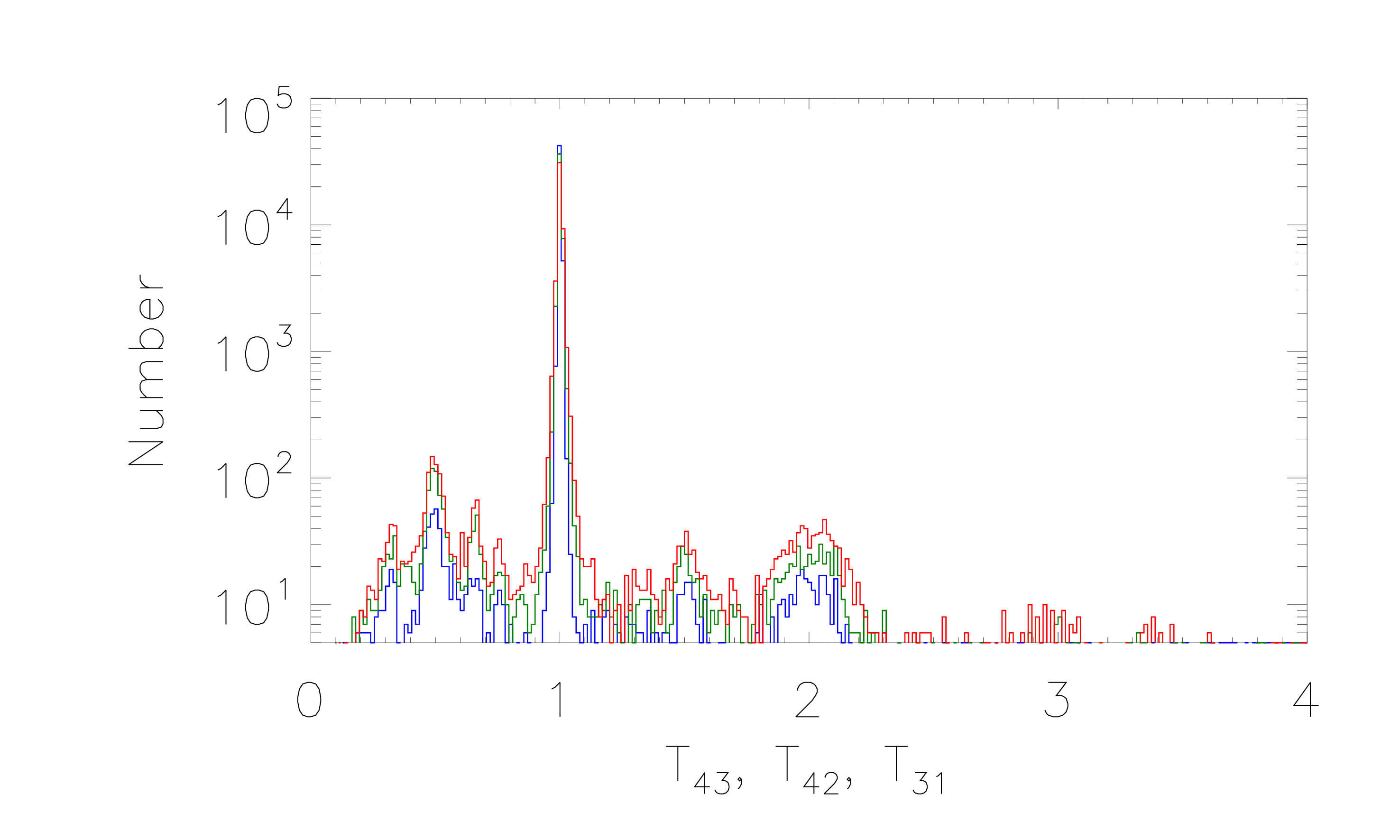}
\caption{Distributions of $T_{43}$ (in blue), $T_{42}$ (in dark green) and $T_{31}$ (in red) of the 50000 artificial light 
curves $LC_{t, obs}$ and the corresponding additional light curves $LC_{43}$, $LC_{42}$, $LC_{31}$. 
}
\label{ppo}
\end{figure}

%%%2nd
	Moreover, it is necessary to check probable effects of the collected model parameters on the $T_{po}$. Here, for the 
parameter pairs [$\tau_1$, $\tau_2$], [$\sigma_{*,1}$, $\sigma_{*,2}$], [$LC_{m1}$, $LC_{m2}$], [$k_{10}$, $k_{20}$] and 
[$\alpha_1$, $\alpha_2$] for the $LC_{t,~1}$ and $LC_{t,~2}$, distributions of $T_{po}$ are compared for the $LC_{t, obs}$ 
with the absolute difference between each parameter pair larger than or smaller than the mean value of the absolute difference, 
such as the compared results between the $LC_{t, obs}$ with $|\tau_1-\tau_2|<300days$ (300days as the mean value of 
$|\tau_1-\tau_2|$ for all the 50000 $LC_{t, obs}$) and the $LC_{t, obs}$ with $|\tau_1-\tau_2|>300days$. The compared 
distributions of $T_{po}$ are shown in Fig.~\ref{mpo}, and there are not different distributions of $T_{po}$, indicating few 
effects of the collected model parameters on the shown results in the right panel of Fig.~\ref{tpo}.

%%%3rd
	Furthermore, based on the 50000 artificial light curves created above, the following procedures are done to check effects 
of different time steps of light curves on our final results. Based on each artificial light curve, three additional light 
curves $LC_{43}$, $LC_{42}$, $LC_{31}$ are created through 3/4, 2/4 and 1/3 of the data points randomly collected. And the 
GLS method is applied to determine the periodicities $T_{p43}$, $T_{p42}$ and $T_{p31}$ in the three additional light curves. 
Then, distributions of ratios of $T_{43}=T_{o}/T_{p43}$ (mean value 1.009), $T_{42}=T_{o}/T_{p42}$ (mean value 1.016) and 
$T_{31}=T_{o}/T_{p31}$ (mean value 1.029) are shown in Fig.~\ref{ppo} for the $LC_{t, obs}$ and the corresponding additional 
light curves $LC_{43}$, $LC_{42}$, $LC_{31}$. There are 97.3\% of the $LC_{43}$ having $0.9<T_{43}<1.1$, 95.1\% of the 
$LC_{43}$ having $0.9<T_{42}<1.1$, and 92.9\% of the $LC_{31}$ having $0.9<T_{31}<1.1$. Therefore, there are slight effects 
of different time steps of the light curves on the final determined periodicities. Meanwhile, considering the mean values 
with $T_{43}<T_{42}<T_{31}$, larger time steps of a light curve should lead to smaller measured periodicity.

%%%4th
	Before ending the subsection, the following four additional points are discussed. First, besides the 50000 artificial 
light curves created above, 50000 another light curves are created with the same values of $\alpha_1=\alpha_2$, and 50000 
another light curves are created with the same values of $k_{10}=k_{20}=0$, and 50000 another light curves are created with 
the time durations very different from those applied above based on light curves of the other quasars from the ZTF. Then, 
based on the re-created light curves, similar results can be confirmed as the results shown in Fig.~\ref{tpo} and in 
Fig.~\ref{mpo} and in Fig.~\ref{ppo}, to support that the collected values of $\alpha$ and $k_0$ and different time durations 
having few effects on our final results, probably indirectly indicating central geometric structures having similar effects on 
obscurations due to orbital rotations related to sub-pc BBHs. Second, upper boundaries $k_{u10}$ and $k_{u20}$ of minimum 
values of $k_{10}<k_{u10}$ and $k_{20}<k_{u20}$ have been collected to be larger than 0.6, leading to part of artificial 
light curves have the measured periodicities $T_p$ with confidence levels smaller than $5\sigma$, due to weaker obscurations 
due to orbital rotations. However, after removing the weaker QPOs, the similar results can be confirmed as the results 
shown in Fig.~\ref{tpo} and in Fig.~\ref{mpo} and in Fig.~\ref{ppo}, not providing clues to change the effects of intrinsic 
AGN variability on optical QPOs. Third, as discussed above, intrinsic AGN variability have apparent and stable effects on 
optical QPOs, indicating there could be probably different periodicities in light curves in different epochs for one assumed 
sub-pc BBH. Although we cannot clearly find the key role to control the variability of the periodicities in different epochs, 
the results could be applied to explain why there were weaker evidence for the periodicity in the known PG 1302-102 if the 
ASAS-SN (All-Sky Automated Survey for SuperNovae) data were combined with the previous LINEAR+CRTS (Lincoln Near-Earth 
Asteroid Research, Catalina Real-time Transient Survey) data discussed in \citet{lg18}, due to probably different measured 
periodicity in the light curves from the ASAS-SN.

\subsection{further discussions}

%%1
	As discussed above, the time information of the ZTF g-band light curve of PG 1411+442 is applied to create the 
simulated light curves, it is necessary to check whether different time information can lead to the same final results. 
Here, light curves of another ten randomly collected SDSS quasars have been collected from ZTF (g-band), and the corresponding 
time information are accepted to create the simulated light curves through the same procedures above. And then, totally 
similar results can be confirmed that about 40\%-45\% ($(N_{po2} + N_{po3} + N_{po4})/50000$) of the 50000 $LC_{t, obs}$ 
have the measured periodicity $T_{o}$ very smaller (at least two times smaller) than the input periodicity $T_p$, due to 
the expected apparent effects of intrinsic AGN variability. Table~1 lists the basic information of the light curves and 
the corresponding results on the simulated results based on the time information of the ZTF light curves of the ten collected 
SDSS quasars. Here, besides the listed results in Table 1, there are no plots on the results, due to totally similar as the 
ones shown in Fig.~\ref{tpo}.

\begin{table*}  
\centering  
	\caption{Basic results through the collected ZTF light curves of the 10 SDSS quasars}   
\begin{tabular}{llllllllllll}
\hline
	obj & N & $T_d$ & mag & $N_{po1}$ & $N_{po2}$ & $N_{po3}$ & $N_{po4}$ & $N_p$ & $T_{43}$ & 
	$T_{42}$ & $T_{31}$\\  
\hline\hline
	PG 1411+442  & 532  & 1992 & 14.99$\pm$0.012 & 29240 & 18336 & 1901 & 523 &   55.9:35.1:3.63:1  &
	1.009 & 1.016 & 1.029 \\
	J02:18:38.90-04:20:41.28 & 284 & 1678 & 18.41$\pm$0.044 & 28244 & 19522 & 1712 & 522 &   54.1:37.4:3.28:1  &
        1.033 & 1.055 & 1.079 \\
	J04:37:21.57+12:58:31.43 & 220 & 1913 & 18.34$\pm$0.040 & 28250 & 19024 & 2113 & 613 &   46.1:31.1:3.45:1  &
        1.036 & 1.059 & 1.083 \\
	J09:44:00.33+28:52:26.75 & 248 & 1890 & 18.00$\pm$0.034 & 30113 & 17697 & 1677 & 513 & 58.7:34.5:3.27:1  &
        1.036 & 1.057 & 1.081\\
	J10:10:44.49+00:43:31.08 & 200 & 1879 & 16.12$\pm$0.019 & 28490 & 18876 & 2051 & 583 &   48.9:32.4:3.52:1 &
	1.037 & 1.063 & 1.084 \\
	J11:27:36.88+24:49:23.52 & 318 & 1915 & 17.44$\pm$0.022 & 29081 & 18460 & 1956 & 503 &  57.8:36.7:3.89:1  &
        1.033 & 1.047 & 1.072 \\
	J12:06:48.31+06:59:12.12 & 361 & 1914 & 17.57$\pm$0.024 & 28931 & 18451 & 2045 & 573 &   50.5:32.2:3.57:1  &
        1.029 & 1.039 & 1.074 \\
        J14:54:34.34+08:03:36.35 & 309 & 1922 & 16.75$\pm$0.018 & 27765 & 19251 & 2361 & 623 &   44.6:30.9:3.79:1  &
        1.035 & 1.043 & 1.075 \\
	J16:27:50.54+47:36:23.39 & 393 & 1921 & 17.62$\pm$0.021 & 28787 & 18605 & 2047 & 561 &  51.3:33.2:3.65:1  &
        1.022 & 1.041 & 1.058 \\
	J21:12:04.84-06:35:35.16 & 244 & 1865 & 17.76$\pm$0.029 & 28386 & 18921 & 2098 & 595 &  47.7:31.8:3.53:1  &
        1.034 & 1.049 & 1.069 \\
	J21:23:47.83-08:28:42.60 & 216 & 1862 & 18.16$\pm$0.031 & 28476 & 18890 & 2048 & 586 &  48.6:32.2:3.49:1  &
        1.038 & 1.066 & 1.079 \\
\hline
\end{tabular}\\
\begin{tablenotes}
\item Notes: The first column shows the name of the collected 10 SDSS quasars in the format of SDSS Jhh:mm:ss.ss$\pm$dd:mm:ss.ss, 
except the PG 1411+442. The second column and the third column show the number of data points and the time duration in 
the collected 1day binned ZTF g-band light curve of each quasar. The fourth column shows the mean magnitude and the mean 
uncertainty of the collected 1day binned ZTF g-band light curve of each quasar. The fifth to eighth column shows the 
determined $N_{po1}$, $N_{po2}$, $N_{po3}$, $N_{po4}$ among the 50000 simulated light curves with the input time information 
from the collected 1day binned ZTF g-band light curve of each quasar. The ninth column shows the determined 
$N_p=N_{po1}:N_{po2}:N_{po3}:N_{po4}$. The last three columns show the determined mean values of $T_{43}=T_{o}/T_{p43}$, 
$T_{42}=T_{o}/T_{p42}$ and $T_{31}=T_{o}/T_{p31}$.
\end{tablenotes}
\end{table*}

%%%2
	Moreover, based on the time information of the ZTF g-band light curves of the collected 10 SDSS quasars, similar 
procedures have been done to check effects of different time steps of the light curves on the final determined periodicities, 
through checking properties of additional light curves $LC_{43}$, $LC_{42}$, $LC_{31}$ created through 3/4, 2/4 and 1/3 of 
the data points randomly collected through the artificial light curves. As the shown results on the mean values of the 
corresponding $T_{43}$, $T_{42}$ and $T_{31}$ in the last three columns in Table 1, it can be re-confirmed slight effects of 
different time steps of the light curves on the final determined periodicities. Here, besides the listed results on mean 
values of $T_{43}$, $T_{42}$ and $T_{31}$ in Table 1 based on the time information of the ZTF g-band light curves of the 
collected 10 SDSS quasars, there are no plots on distribution of $T_{43}$, $T_{42}$ and $T_{31}$, due to totally similar 
as the ones shown in Fig.~\ref{ppo}. Meanwhile, when additional light curves $LC_{42}$ and $LC_{31}$ are 
created through 2/4 and 1/3 of the data points randomly collected from the artificial light curves, the quality (the value 
of time duration divided by the number of data points) of the $LC_{42}$ and $LC_{31}$ should be similar as that of common 
light curves collected from the CSS (Catalina Sky survey) \citep{dd09}. Therefore, there are no further discussions on 
simulated results through time information of CSS light curves.

%%%3
	Furthermore, due to the discussions above on the sub-pc BBHs, it is interesting to check the probable beaming 
effects by probably relativistic orbital rotations on observational photometric properties. Based on discussed BBH model 
in \citet{eb12}, the expected space separation of the central two BH accreting systems can be estimated as 
\begin{equation}
	S_{BBH}~\sim~0.432M_{8}(\frac{T_{qpo}}{2652M_{8}})^{2/3}~pc
\end{equation}
with $M_8$ as the total BH mass in units of $10^8{\rm M_\odot}$ and $T_{qpo}$ as the orbital periodicity in units of years. 
Meanwhile, assumed a simple circular orbital, the sum $V$ of the absolute orbital velocities $|V_1|$ and $|V_2|$ of the two 
BHs can be expected as
\begin{equation}
\begin{split}
V &= |V_1|+|V_2| = \sqrt{\frac{GM_{8}}{S_{BBH}}} \\
	&=13814{\rm km/s}~\times~(\frac{M_8}{T_{qpo}})^{1/3}
\end{split}
\end{equation}
with $G$ as the gravitational constant. After simply considering $M_8\sim10$ and $T_{qpo}\sim2.74$years (mean values of the 
sub-pc BBH system candidates listed in \citealt{gd15}), the determined Lorentz factor is only about 
$\frac{1}{\sqrt{1-(V/c)^2}}\sim1.0025$ ($c$ as the light speed), indicating very weak beaming effects by orbital rotation 
on the observational photometric properties. Therefore, there are no further discussions on beaming 
effects on our discussed results.

	Before ending the section, we should note that the applied CAR process and the piecewise functions are not completely 
accurate but only mathematics-oriented approach for intrinsic AGN variability after considering the obscurations related to 
orbital motions in sub-pc BBH systems. Therefore, in order to test and confirm our results reported and discussed in this 
manuscript, it will be our objective in the near future to test whether are there different periodicities in different epochs 
from those of the optical QPOs reported in the literature.

\section{Main summary and Conclusions}
	Our main summary and conclusions are as follows.
\begin{itemize}
\item Considering intrinsic AGN variability described by the CAR process, combining with orbital rotations expected obscurations 
	described by piecewise functions, effects of intrinsic AGN variability on optical QPOs related to sub-pc BBHs can be 
	checked.
\item The measured periodicities $T_{o}$ are not well consistent with the input orbital periodicity $T_{p}$, only about 41.52\% 
	of the artificial light curves have $T_{o}\sim T_p$. However, about 36.67\% of the artificial light curves have 
	$T_{o}\sim T_p/2$, about 3.8\% of the artificial light curves have $T_{o}\sim T_p/3$, and about 1.05\% of the 
	artificial light curves have $T_{o}<T_p/4$. After considering effects of the collected model parameters, total similar 
	results can be confirmed, indicating strong and stable effects of the intrinsic AGN variability on optical QPOs. 
\item Moreover, based on the light curves with data points randomly collected from the artificial light curves, larger time 
	steps of light curves have tiny effects on the measured optical periodicities, and can probably lead to a bit smaller 
	measured optical periodicities.
\item It will be our objective in the near future to check whether the reported optical QPOs in the literature should have 
	different periodicities detected through light curves in different epochs from different sky survey projects.
\end{itemize}

\section*{acknowledgments}
Zhang gratefully acknowledge the anonymous referee for giving us constructive comments and suggestions to 
greatly improve the paper. Zhang gratefully thanks the kind financial support from GuangXi University and the kind grant 
support from NSFC-12173020 and NSFC-12373014. This paper has made use of the data from the ZTF \url{https://www.ztf.caltech.edu}.

%\section*{Data Availability}
%The data underlying this article will be shared on reasonable request to the corresponding author
%(\href{mailto:xgzhang@gxu.edu.cn}{xgzhang@gxu.edu.cn}).

\label{lastpage}
\end{document}